\documentclass[sigconf]{acmart}

\setcopyright{none}
\copyrightyear{2019}
\acmYear{2019}
\acmConference{ACSAC'19}
\acmBooktitle{}

\usepackage{graphicx} 
\usepackage{subfigure} 

\usepackage{natbib}

\usepackage{algorithm}
\usepackage{algorithmic}
\usepackage{mathtools}
\usepackage{blkarray, bigstrut} 
\usepackage{float}
\usepackage{multirow}
\usepackage{colortbl}
\usepackage{hyperref}

\usepackage{xspace}
\usepackage{xcolor}
\usepackage[export]{adjustbox}

\usepackage{mdframed}

\usepackage{url}

\usepackage{breakurl}

\definecolor{high}{HTML}{CC92CB}
\definecolor{med}{HTML}{8AC667}
\definecolor{low}{HTML}{8DBAA1}
\definecolor{veryhigh}{HTML}{936BAF}
\definecolor{verylow}{HTML}{CEE0D6}
\definecolor{mediumseagreen}{HTML}{3CB371}

\setcitestyle{numbers}
\setcitestyle{square}

\usepackage[
]{todonotes}

\def\etal{{et al.}\xspace}

\setcopyright{acmcopyright}

\begin{document}

\pagestyle{plain}  
\pagestyle{empty}  

\copyrightyear{2019}
\acmYear{2019}
\acmConference[ACSAC '19]{2019 Annual Computer Security Applications Conference}{December 9--13, 2019}{San Juan, PR, USA}
\acmBooktitle{2019 Annual Computer Security Applications Conference (ACSAC '19), December 9--13, 2019, San Juan, PR, USA}
\acmPrice{15.00}
\acmDOI{10.1145/3359789.3359835}
\acmISBN{978-1-4503-7628-0/19/12}

\newcommand{\lastline}{Local Vendor\xspace}
\newcommand{\cmmnt}[1]{\ignorespaces}
\newcommand{\cj}[1]{\textcolor{red}{\{CJ: #1\}}}
\newcommand{\cs}[1]{\textcolor{red}{\{CS: #1\}}}
\newcommand{\deepnet}{Neurlux\xspace}
\newcommand{\emberdataset}{\emph{EmberDataset}\xspace}
\newcommand{\wilddataset}{\emph{VendorDataset}\xspace}
\newcommand{\cuckooreport}{\emph{CuckooSandbox}\xspace}
\newcommand{\lastlinereport}{\emph{VendorSandbox}\xspace}
\newcommand{\rawmodel}{\emph{Raw Model}\xspace}
\newcommand{\countsmodel}{\emph{Counts Model}\xspace}
\newcommand{\documentmodel}{\deepnet}
\newcommand{\individualmodel}{\emph{Individual Model}\xspace}
\newcommand{\ensemblemodel}{\emph{Ensemble Model}\xspace}
\newcommand{\maldy}{\emph{MalDy}\xspace}

\title{\deepnet: Dynamic Malware Analysis Without Feature Engineering}

\author{Chani Jindal}
\affiliation{%
  \institution{\hbox{University of California, Santa Barbara}}
}
\affiliation{%
      \institution{Appfolio}
}
\email{chanijindal@ucsb.edu}

\author{Christopher Salls}
\affiliation{%
  \institution{\hbox{University of California, Santa Barbara}}
}
\email{salls@cs.ucsb.edu}

\author{Hojjat Aghakhani}
\affiliation{%
  \institution{\hbox{University of California, Santa Barbara}}
}
\email{hojjat@cs.ucsb.edu}

\author{Keith Long}
\affiliation{%
  \institution{\hbox{University of California, Santa Barbara}}
}
\email{klong@ucsb.edu}

\author{Christopher Kruegel}
\affiliation{%
  \institution{\hbox{University of California, Santa Barbara}}
}
\affiliation{%
      \institution{Lastline}
}
\email{chris@cs.ucsb.edu}

\author{Giovanni Vigna}
\affiliation{%
  \institution{\hbox{University of California, Santa Barbara}}
}
\affiliation{%
      \institution{Lastline}
}
\email{vigna@cs.ucsb.edu}

\begin{abstract}
Malware detection plays a vital role in computer security. Modern machine learning approaches have been centered around domain knowledge for extracting malicious features. However, many potential features can be used, and it is time consuming and difficult to manually identify the best features, especially given the diverse nature of malware. 

In this paper, we propose \deepnet, a neural network for malware detection. \deepnet does not rely on any feature engineering, rather it learns automatically from dynamic analysis reports that detail behavioral information. Our model borrows ideas from the field of document classification, using word sequences present in the reports to predict if a report is from a malicious binary or not. We investigate the learned features of our model and show which components of the reports it tends to give the highest importance. Then, we evaluate our approach on two different datasets and report formats, showing that \deepnet improves on the state of the art and can effectively learn from the dynamic analysis reports. Furthermore, we show that our approach is portable to other malware analysis environments and generalizes to different datasets.
\end{abstract}



\maketitle

\section{Introduction}
As malware becomes more sophisticated, malware analysis needs to evolve as well. Traditionally, most anti-malware software uses signature-based detection, which cross-references executable files with a list of known malware signatures. However, this approach has limitations, since any changes to malware can change the signature, so new releases of the same malware can often evade signature-based detection by encrypting, obfuscating, packing, or recompiling the original sample.
VirusTotal reports that over 680,000 new samples are analyzed per day~\cite{vt-comparative-analyses}, of which potentially a significant number of samples are just re-packed versions of previously seen samples, as Brosch \etal~\cite{brosch2006runtime} observed more than 50\% of new malware are simply re-packed versions of existing malware.

In recent years, the need for techniques that generalize to previously unseen malware samples has led to detection approaches that utilize machine learning techniques \cite{ucci2017survey, raff2018malware, raff2017learning}.
Malware analysis can be broadly divided into two categories: code (static) analysis and behavioral (dynamic) analysis. Both static and dynamic analysis have their advantages and disadvantages.
Although dynamic analysis provides a clear picture of the executable behavior, it faces some problems in practice.
For example, dynamic analysis of untrusted code requires a virtual machine that replicates the target host, which requires a substantial amount of computing resources.
Besides, malware may not exhibit its malicious behavior, or the virtualized environment may not reflect the environment targeted by the malware~\cite{rossow2012prudent, raffetseder2007detecting, lindorfer2011detecting, garfinkel2007compatibility}.

To avoid such limitations, some related work relies only on features extracted from static analysis to achieve rapid detection for a large number of malware samples.
However, various encryption and obfuscation techniques can be employed to hinder static analysis~\cite{moser2007limits, perdisci2008mcboost}.
This becomes a more severe problem for static malware detectors, since packing is also in widespread use in benign samples today. samples~\cite{rahbarinia2017exploring}. Although dynamic analysis is shown to be susceptible to evasion techniques, run-time behavior is hard to obfuscate. Dynamically analyzing a binary gives the ability to unpack and record its interactions with the OS which it an attractive choice for malware analysis.


 Regardless of the use of static analysis or dynamic analysis, most machine-learning based malware detectors rely heavily on relevant domain knowledge~\cite{kolbitsch2009effective, sgandurra2016automated, kharaz2016unveil}. These approaches often rely on features that are investigated manually by malware experts, which requires a vast amount of feature engineering.
For example, Kolbitsch \etal~\cite{kolbitsch2009effective} captured the behavior graphs of PE executables in specific features designed for this purpose.
 Malware is continually being created, updated, and changed, which can make the original well-designed features not applicable to newer malware or different malware families. In this case, the costly feature engineering work has to be refined continuously. Hence, it is crucial to find a way toreduce the cost of artificial feature engineering to extract usefulinformation from raw data.
 
 There has recently been some work on deep learning based malware classification which does not require feature engineering. 
 However, existing deep learning approaches do not leverage the information from already-available dynamic analysis systems, instead tending to pick one type of dynamic feature \cite{kolosnjaji2016deep} or use static features \cite{grosse2017adversarial}.
 These solutions miss out on the complete information concerning what actions are taken by each sample.

In this paper we propose Neurlux, a system that uses neural networks to analyze dynamic analysis reports. Services such as Cuckoo \cite{oktavianto2013cuckoo} provide a detailed dynamic analysis of an executable by tracing it in a sandbox. This analysis contains information, such as network activity, changes to the registry, file actions, and more. We use such reports as the basis for our analysis. That is, given a dynamic analysis report, we want to be able to predict whether or not the report is for a malware sample or a benign executable.

Our intuition is that we can treat these reports as documents. With this intuition, we present \deepnet, a neural network which learns and operates on the (cleaned) dynamic analysis report without needing any feature engineering. 
\deepnet borrows concepts from the field of document classification, treating the report as a sequence of words, which make sequences of sentences, to create a useful model.
\deepnet intends to replace expensive hand-crafted heuristics with a neural network that learns these behavioral artifacts or heuristics.

To check if our method is biased to a particular report format (i.e., sandbox), we included in our evaluation two different sandboxes, the Cuckoo sandbox~\cite{oktavianto2013cuckoo}, \cuckooreport and a commercial anti-malware vendor's sandbox, which we will refer to as \lastlinereport.
In addition, we used two different datasets, one provided by the commercial anti-malware vendor, \wilddataset along with the labeled benchmark dataset EMBER~\cite{anderson2018ember}, \emberdataset.

To show that \deepnet does better than feature engineering approaches, we implement and compare against three such techniques which are discussed later. Furthermore, we implement and compare against MalDy, a model proposed by Karbab \etal~\cite{karbab2019maldy}, as a baseline. MalDy formalizes the behavioral (dynamic) report into a bag of words (BoW) where the features are words from the report.

In summary, we make the following contributions:
\begin{itemize}
    \item We propose \deepnet, an approach which leverages document classification concepts to detect malware based on the behavioral (dynamic) report generated by a sandbox without the need for feature engineering.
    The only preprocessing step is cleaning the reports to extract words, upon which our model learns relevant sequences of words which can aid its prediction. \deepnet shows high accuracy achieving 96.8\% testing accuracy, in our K-fold validation.
    \item We create and test several approches for malware classification on dynamic analysis reports, including novel methods such as, a Stacking Ensemble for Integrated Features, and a Feature Counts model. We compared with these, showing \deepnet outperforms approaches with feature engineering. 
    \item We assess the generalization ability of \deepnet by testing it against a new dataset and also a new report format, i.e., generated by a new sandbox and show that it generalizes better than the methods we evaluated against. 
    \item The source code and dataset of executables will be released on github.
\end{itemize}

\section{Background}
\label{sec:background}
Some related work adopted Natural Language Processing techniques for malware classification such as MalDy~\cite{karbab2019maldy}, which formalizes the behavioral report of a sample into a bag of words.
In this section, we explain such techniques that we exploited to build \deepnet and other models as a baseline for comparison to \deepnet. 
\subsection{Word Embeddings}
\label{sec:wordembedding}
Word embeddings are translations from words to vectors that aim to give words with similar meaning corresponding vectors that are close in the feature space.
Word embeddings are frequently found as the first data processing layer in a deep learning model that processes words \cite{trendsinnlp}. This is because grouping vectors by meaning gives a deep learning model an initial correlation of words. These embeddings are frequently pre-trained or based on another model such as word2vec \cite{mikolov2013distributed}. However, in our case, we do not use a pre-trained embedding as the similarities of "words" (i.e., file paths, mutexes, etc.) differ from ordinary English.

\subsection{Embedding Visualization}
Dimensionality reduction methods are used to convert high dimensional data into lower dimensional data. We can use dimensionality reduction to convert the data into two dimensions, allowing us to show the distribution of the data in a scatter plot. To do this, we choose to use t-Distributed Stochastic Neighbor Embedding (t-SNE) \cite{maaten2008visualizing}, a technique for visualization of similarity data. t-SNE preserves the local structure of the data and some global structure, such as clusters, while reducing the dimensionality.

\subsection{CNN for text classification}
Convolutional neural networks (CNN) have recently shown to be very useful in text classification. A typical model for this represents each input as a series of n sequences, where each sequence is a d-dimensional vector; thus input is a feature map of dimensions $n \times d$.
The model starts by mapping words to vectors, as discussed in Section~\ref{sec:wordembedding}. Then, convolutional layers are used for representation learning from sliding $k-grams$. 

To extract higher level features from input vectors, a CNN applies filters of $R^{k \times d}$ on an input of length $n$,  $\{x_1, x_2, x_3, x_4, x_i .... x_n\}$. After applying a filter of size $k$ we have,  $\{x_{1:k}, x_{2:k+1}, x_{3:k+2}, .... x_{n-k+1:n}\}$. Embeddings
for $x_i$, $i < 1$ or $i > n$, are zero padded. For each window, $x_{i:i + k - 1}$,  a feature $p_i$ is generated which is then fed
into ReLU non-linearity. 
\begin{equation} \label{eq:20}
p_i = f(W.k^T + b)
\end{equation}
Where $b \in R$ is a bias term, $f$ is a non-linear activation function, such as the hyperbolic tangent, and $W.k$ is the weights for filter $k$. Applying filter $k$ to all windows results in the feature map.

Max pooling sub-samples the input by applying a max operation on each sample. It extracts the most salient n-gram features across the sentence in a translation-invariant manner. The extracted feature can be added anywhere in the final sentence representation.

In practice we use multiple window sizes and multiple convolutional layers in parallel. A combination of convolution layers followed by max pooling is often used to create deep CNN networks. Sequential convolutions can improve the sentence mining process by capturing an abstract representation which is also semantically rich.

\subsection{LSTM/ BiLSTM}
Recursive Neural Networks (RNN) have gained popularity with text classification due to their ability to preserve sequence information over time. LSTM networks \cite{hochreiter1997long} overcome the vanishing gradient problem of RNN \cite{hochreiter1998vanishing}. LSTM networks use an adaptive gating, which regulates the flow of information from the previous state and the extracted features of the current data input. For an input sequence with $n$ entries: $x_1, x_2,..., x_n$, an LSTM network processes it word by word. Then, it uses the following equations to update the memory $p_t$ and hidden state $h_t$ at time-step t:

\begin{equation} \label{eq:2}
\begin{bmatrix}i_t \\ f_t \\ o_t \\ q_t\end{bmatrix} = 
\begin{bmatrix}\sigma \\ \sigma \\ \sigma \\ tanh \end{bmatrix} W.\big[ h_{t-1}, x_t \big]
\end{equation}

\begin{equation}\label{eq:3}
    p_t = f_t \ast p_{t-1} + i_t \ast q_t
\end{equation}

\begin{equation}\label{eq:4}
    h_t = o_t \ast tanh(p_t)
\end{equation}

where $x_t$ is the input at time-step $t$, $i_t$ is the input gate activation, $f_t$ is the forget gate activation, and $o_t$ is the output gate activation. All gates are generated by a sigmoid function, $\sigma$ over the ensemble of input $x_t$ and the preceding hidden state $h_{t-1}$. 

A BiLSTM network extends the unidirectional LSTM by initiating a second hidden layer. In this layer, the hidden-to-hidden connections can flow in reverse temporal order. Therefore, the model holds information from both the past and the future. The output of $j^{th}$ word can be represented as:
\begin{equation}\label{eq:5}
    h_j = \big[ \overrightarrow{h_j}  \bigoplus \overleftarrow{h_j} \big]
\end{equation}

\subsection{Attention}
\label{sec:attention}
It is evident that not all words contribute equally to the malicious or benign attributes of dynamic behavior. Hence, at the word level, an attention mechanism \cite{vaswani2017attention} can be used to extract malicious features/words that are important to the behavior classification. Finally, we aggregate the representations of those malicious features to form the sentence representation.

Let $H \in R^{d\times n}$ be a matrix of hidden vectors $[h_1, h_2,...,h_n]$ that the LSTM network produced, where $d$ is the size of the hidden layers, and $n$ is the number of words in a sentence. Let $h_{it} \in H$ represent a hidden state. The first step is to feed $h_{it}$ through a single-layer Perceptron network to get $u_{it}$ as its hidden representation:

\begin{equation}\label{eq:7}
    u_{it} = tanh(W_w h_{it} + b_w).
\end{equation}

The second step is to initialize the context vector $u_w$ randomly. Then, the importance of the word as the similarity of $u_{it}$ with a word-level context vector $u_w$ is measured. This gives a normalized importance weight vector $\alpha$ through a softmax function. The Context vector $u_w$ is learned during the training process. $\alpha$ measures the importance of $i^{th}$ word for malicious behavior. 

\begin{equation}\label{eq:8}
    \alpha_{it} = \frac{exp(u_{it}^\top u_w)}{\sum_t exp(u_{it}^\top u_w)}
\end{equation}

In the end the sentence can be represented as the weighted hidden vector $r$:
\begin{equation}\label{eq:9}
    r = H\alpha^T.
\end{equation}




\begin{figure*}
\begin{center}
\centerline{\includegraphics[width=\linewidth]{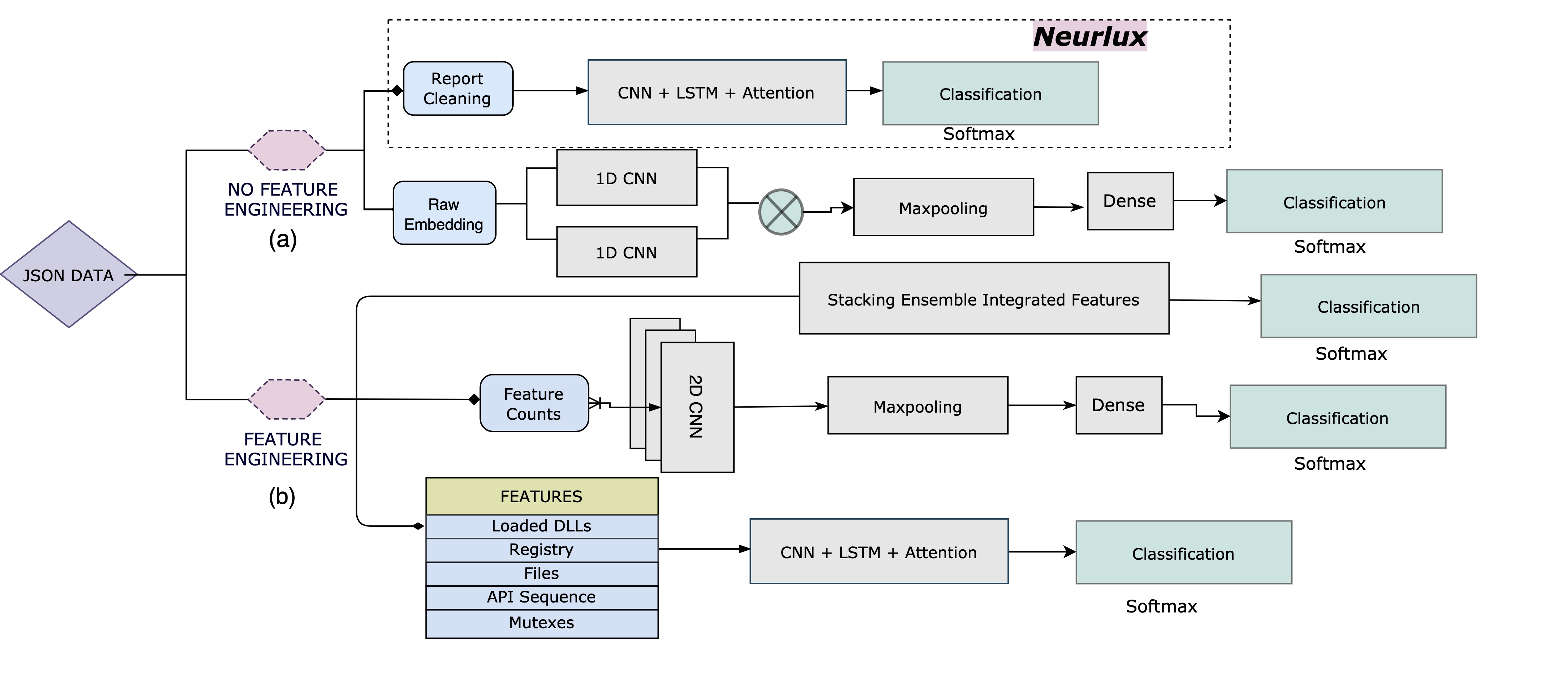}}
\caption{Overview of Models}
\label{fig:overview}
\end{center}
\vskip -0.2in
\end{figure*}

\section{Approach}
\begin{figure*}
\begin{center}
\centerline{\includegraphics[width=.7\linewidth]{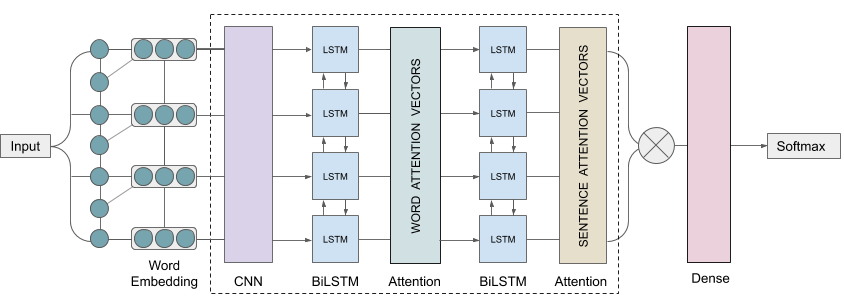}}
\caption{CNN-BiLSTM-Attention model for document classification}
\label{fig:bilstm}
\end{center}
\end{figure*}

\label{sec:noengineering}

In this section, we describe our proposed method, \deepnet, a model which treats the dynamic behavior classification problem in a way similar to a document classification problem. The steps of this approach are as follows:

\begin{itemize}
    \item \textbf{Data cleaning:}
    We want to treat the reports as a document classification task, so the first step is to clean the JSON formatted reports so that it is structured less like a JSON document and more like sequences of words, which makes up sequences of sentences.
    To do this, we first remove special characters, such as the brackets which are part of the JSON structure. Then the document is tokenized to extract words, of which the top 10,000 most common words are converted into numerical sequences.
    
    \item \textbf{Data Formatting:} The na\"ive method of converting words to vectors assigns each word with a one-hot vector. This vector would be all zeros except one unique index for each word. This kind of word representation can lead to substantial data sparsity and usually means that we may need more data to train statistical models successfully. This can be fixed by continuous vector space representation of words. To be more specific, we want semantically similar words to be mapped to nearby points, thus making the representation encode useful information about the words' actual meaning.
    
    Therefore, we use trainable word embeddings in \deepnet, which can have the property that similar words have similar vectors as described in Section~\ref{sec:wordembedding}. This way, the model can cluster words based on their usage patterns and they can provide more meaningful inputs to the later layers in the model.
    
    \item \textbf{Model:} We use the combination of CNN, BiLSTM networks, and Attention networks to create a model that understands the hidden lexical patterns in the malicious and benign reports. This model is designed using concepts from document classification. For example, an important idea is that not all words in a sentence are equally important, so it uses the attention mechanism to recognize and extract important words \cite{yang2016hierarchical}.
    Another aspect is that context is important for understanding the meaning of words, therefore we use BiLSTM to give context to the model.
    This model is described in detail in the next section (~\ref{sec:cbia_model}).
\end{itemize}

\subsubsection{ \textbf{CNN + BiLSTM + Attention}}
\label{sec:cbia_model}
 Inspired by the previous document classification methods, we create the model illustrated in  Figure~\ref{fig:bilstm}, which consists of a convolutional neural network layer (CNN), and two pairs of bi-directional long short term memory network (BiLSTM) and attention layers. 
 Convolutional neural networks (CNN) extract local and deep features from the input text. Then we obtain the high-level representation from the bidirectional LSTM network by using the hidden units from both the forward and backward LSTM. 
The CNN and LSTM combination is useful to extract local higher level features, which the LSTM can then find temporal relationships between \cite{zhou2015c}. 
 The two pairs of BiLSTM and Attention are inspired by hierarchical attention networks \cite{yang2016hierarchical}.
 The input is a trainable word embedding of dimension 256 to allow the model to cluster similar words as it learns.
 Each of these components is described thoroughly in the Section~\ref{sec:background}.

As described before, different parts of the report have different importance for determining the overall malicious behavior of a binary. For example, some parts of a registry key can be decisive, while others are irrelevant. Suppose we have the sentence attention score $A_i$
for each sentence $s_i \in x$, and the word attention
score $a_{i, j}$, for each word $w_{i,j} \in s_i$; both scores are normalized which satisfy the following
equations, 
\begin{equation}\label{eq:10}
    \sum_i{A_i} = 1 \hspace{.5cm} and \hspace{.5cm} \sum_j{a_{i,j}} = 1.
\end{equation}

The sentence attention measures which sentence is more important for the overall behavior while the word attention captures behavior signals such
as the behavior words in each sentence. Therefore, the document representation r for document x is calculated as follows,
\begin{equation}\label{eq:11}
    r = \sum_i{\bigg[ A_i . \sum_j \bigg( a_{i,j}.h_{i,j}\bigg)\bigg]}.
\end{equation}

Finally, \deepnet outputs a classification decision as a score from 0-1 where 0 is benign, and 1 is malicious.


\section{Comparison Methods}
\label{sec:comparisons}
In this section, we describe various models with which we compare. We compare with a previous state of the art method and with a couple approaches which involve feature engineering to check if we can actually do better than feature engineering approaches. An overview of the approaches we compare against are shown in Table~\ref{tab:modelrecap}.

\begin{table}[tbp]

\centering
\caption{Overview of the models that we create and compare against \deepnet.}
\label{tab:modelrecap}
\resizebox{\linewidth}{!}{%
\begin{tabular}{l|l|l}
Model & 
\begin{tabular}[c]{@{}l@{}}Feature \\ Engineering \end{tabular} &
Description \\ \hline
\countsmodel & Yes & Counts of each feature \\ \hline
\individualmodel & Yes & 
\begin{tabular}[c]{@{}l@{}}Document classification on \\ individual extracted features \end{tabular} \\ \hline
\ensemblemodel & Yes & Ensemble of individual features \\ \hline
\maldy & No & 
\begin{tabular}[c]{@{}l@{}}State of the art model for  \\ report classification from \cite{karbab2019maldy} \end{tabular} \\ \hline
\rawmodel & No & Model trained on raw bytes \\ \hline
\textbf{\emph{\deepnet}} & \textbf{No} & 
\begin{tabular}[c]{@{}l@{}}\textbf{Document classification on} \\ \textbf{whole report} \end{tabular} \\
\end{tabular}}
\end{table}

\subsection{Comparison With a State-of-the-Art Model}
\label{sec:baseline}
We used the method described in \maldy \cite{karbab2019maldy}, as a model for comparison. Their approach is to preprocess sandbox reports with standard Natural Language Processing (NLP) techniques and then create an ensemble supervised machine learning (ML) model from a multitude of different ML algorithms. They attempt to formalize the behavioral reports in a way agnostic to the execution environment. This is done on both Win32 and Android. They argue that the key to their success is using their bag of words (BOW) model with Common N-Grams (CNG). CNG effectively computes the contiguous sequences of $n$ items where $n$ is an adjustable hyper-parameter. Instead of using single words (1-grams), using n-grams aids in finding distinct features. Once the reports are in a list of n-gram strings, they carry out two different vectorization approaches: TF-IDF and Feature Hashing. Feature Hashing creates fixed length feature vectors from sparse input n-grams. A hash is taken of each n-gram, and if the value is found within the table, it is incremented; otherwise, a value of 1 is added to the table. This process creates probabilistically unique vectors, given that the hash bucket size is sufficiently large. These vectors are subsequently fed into the ML models. We implement and use their best performing model as a comparison. In our evaluation (Section~\ref{sec:evaluation}), this model reaches a plateau with 89.23\% accuracy and an F-1 score of 88.5\%.

A weakness of the BOW approach used in \maldy is that it does not take into account the context, just the frequencies with which words appear \cite{paltoglou2013more}.

\subsection{Raw JSON Data}
A more basic deep learning approach is to learn from raw bytes, treating it as an image classification problem. Although the structure of the input data is defined, the placement of different string objects within the file is not ordered. To best capture such high-level location invariance, we choose to use a convolution network architecture. Combining the convolutional activations with a global max-pooling, followed by fully-connected layers allows this model to produce its activation regardless of the location of the detected features.

The \rawmodel was inspired by an earlier approach on byte classification \cite{raff2018malware}. First, we clean the document, removing special characters.
Then the bytes are extracted as integer values then padded to fix length to form a vector $x$ of $d$ elements.
This ensures that regardless of the length of the input file, the input vector provided to the network has a fixed dimensionality. Each byte $x_{j}$ is then embedded as a vector $z_{j} = \phi(x_{j})$ of eight elements (the network learns a fixed mapping during training). This amounts to encoding $x$ as a matrix \(Z \epsilon R^{[d\times8]}\). Figure~\ref{fig:overview}$(a)$ shows an outline of the model used for raw JSON data binary classification. Then, it goes through the convolutional layers to eventually produce a classification between 0 and 1.

\subsection{Features for Engineering Approaches}
\label{sec:featureengineering}
For the feature engineering approach comparisons, we begin by categorizing the six main categories of features available in the reports. These features are described in more detail below.

\begin{itemize}
\item \textbf{API Sequence Calls.} The reports typically include all system calls and their arguments stored in a representation tailored explicitly to behavior-based analysis. Much of the past work on behavior analysis has focused on using API call sequences for malware classification \cite{shankarapani2011malware, peiravian2013machine, sami2010malware}. 

\item \textbf{Mutexes.} Mutexes control the simultaneous access of the system resources. They are used by malware creators to avoid infecting a system more than once, and coordinate between processes \cite{malwaremutex}.

\item \textbf{File System Changes.} The interaction of a malware sample with the host file system might be a good indicator to determine malicious behavior. We consider all the important file operations such as create, read, write, modify, delete, etc. 

\item \textbf{Registry Changes.} 
The registry is a core part of Windows and contains a plethora of raw data. Registry keys can reveal much information about the system, but the true challenge is in unraveling which modifications to the registry are malicious and which are legitimate. The registry also represents a fundamental tool to hook into the operating system to gain persistence. Discovering what keys are queried, created, deleted, and modified can shed light on many significant characteristics of a sample.

\item \textbf{Loaded DLLs.} The reports contain the shared library code loaded by a program. Nearly every executable program imports DLLs during execution. These DLLs can give insights into the types of APIs used by the program.

\end{itemize}
Figure~\ref{fig:cuckoo_report} is an example of a \cuckooreport report for a malicious sample that shows the various behavioral features cuckoo identifies. We obtain 28 such different features from \cuckooreport and 43 features from \lastlinereport. These features are mapped based on semantic similarities and divided into 6 main behavioral groups as described above. The following sections give an exhaustive description of the various feature engineering techniques used. They are shown in Figure~\ref{fig:overview}$(a)$.


\subsection{Feature Counts Model}
\label{sec:features_count}
In this section, we discuss an approach to use a neural network on shallow numerical features. Numerical features here are simply the counts of each event that was recorded, e.g., number of registry reads, number of file writes, etc. The first step is to parse the reports and extract all the available features. The number of features extracted differs due to the structural differences in dynamic analysis reports collected from \cuckooreport and \lastlinereport. Each report lists features according to the parent process and child processes (any process that was either spawned by or tampered with by the primary process). Each process has its own set of individual features. Since each executable can contain one or more processes, the final representation of input features per sample will be: 
\begin{equation}\label{eq:19}
\begin{split}
    &\mathbf{S} = processes \times features \\
  & \text{which expands to} \\
  & \small{\mathbf{S} =
  \begin{blockarray}{*{4}{c} l}
    \begin{block}{*{4}{c} l}
      reg\_read & file\_write & \cdots & mutex & \\
    \end{block}
    \begin{block}{[*{4}{c}]>{\footnotesize}l<{}}
      a_{1,1} & a_{1,2} & \cdots & a_{1,n} \bigstrut[t]& 1 \\
      a_{2,1} & a_{2,2} & \cdots & a_{2,n} & 2 \\
      \vdots & \vdots & \ddots & \vdots & \vdots \\
      a_{m,1} & a_{m,2}  & \cdots & a_{m,n} & n \\
    \end{block}
  \end{blockarray}
  } \\
& \footnotesize{\text{(Columns are features, rows are processes)}}
\end{split}
\end{equation}

The data representation is similar to that of a gray-scale image; therefore, a 2D CNN can be used for training on this dataset. We use an 8-layer deep CNN model inspired by Simonyan et al. \cite{simonyan2014very}. The model consists of 8 convolution layers and 2 fully-connected layers. Every convolutional filter has a kernel size of 3, 4, or 5 with a stride of 1 and pooling region of 3x3 without overlap. A pooling function is applied to each feature map to induce a fixed-length vector. These fixed-length, top-level feature vectors generated from filter maps are then fed through a softmax function to generate the final classification. Figure~\ref{fig:overview} gives an overview of this model. 

\subsection{Text-Based Individual Feature Model}
\label{sec:text_classify}
Each analysis report is a collection of statements, and each statement is a sequence of words. We believe that these sequences can give a more granular description of the actual events, compared to the features count method discussed in the previous section. The assumption for the text classification approach described in this section is that the difference between malicious and benign behavior of binaries could be translated into sequences present in the reports. In other words, the sequence of actions better represents if a binary is malicious or not than merely the number of actions. Additionally, we were looking for a feature representation (sequences of words) that uses an automatic feature extraction without the intervention of a security expert.

The input generation process can be divided into four steps. This process is performed iteratively for all six feature groups. 

\begin{itemize}
    \item \textbf{Feature Selection:} Different features are selected from the feature pool based on the top six behavioral groups discussed earlier in this section.
    \item \textbf{Data cleaning:} Similar to our method for \deepnet, we need to remove special characters and perform tokenization to extract numerical sequences. 
    \item \textbf{Data Formatting:} As discussed previously, we want to have a continuous vector space representation of words, with semantically similar words mapped to a nearby point. So once again we use a trainable word embedding. 
    \item \textbf{Model training:} We use the combination of CNN, BiLSTM, and Attention networks to create a model that understands the hidden lexical patterns in the malicious and benign reports. This model is described in detail in Section~\ref{sec:cbia_model}
\end{itemize}

\subsection{Integrated Features using Ensemble}
When the neural network, described in Section~\ref{sec:cbia_model}, is trained on individual behavioral feature types (such as mutexes or api calls), it exhibits a high variance depending on the feature. This variance can be attributed to the importance and contribution of each feature extracted from the reports. Therefore, in this section, we describe a way to use ensemble learning to combine multiple models to get a low variance and better predictive performance than any single constituent algorithm alone. Specifically, we use a stacking ensemble which uses one large multi-headed neural network and learns how to best integrate predictions from each input sub-network.

We start with five separately trained models from the previous section. These models are trained on the five most important behavior traits observed in the reports; namely, API Sequences, Mutex Operations, File Operations, Registry Operations, and DLLs Loaded. Each model uses the text-based feature classification method explained in Section~\ref{sec:text_classify}. The outputs from all the sub-models are concatenated, and provided to a fully connected layer that acts as a meta-learner, and makes its probabilistic predictions. The sub-models are not trainable; therefore, their weights are not changed during the training, and only the weights of new hidden and output layers are updated.

Merging the outputs from multiple neural networks adds a bias that in turn counters the variance of a single trained neural network model. As a result, the final predictions are less sensitive to the specifics of training data and are more generalized.

\begin{table*}[!ht]
\centering
\begin{tabular}{|l|l|l|l|l|l|}
\multicolumn{6}{c}{\textit{Malware Sample 1}} \\
\hline
\cellcolor{verylow} windows  & \cellcolor{verylow} nt  & \cellcolor{verylow} terminal  & \cellcolor{verylow} services  & \cellcolor{verylow} maxdiscon...  & \cellcolor{verylow} microsoft \\ \hline 
\cellcolor{verylow} windows  & \cellcolor{low} defender  & \cellcolor{verylow} spynet  & \cellcolor{verylow} spynetreporting  & \cellcolor{verylow} microsoft  & \cellcolor{verylow} windows \\ \hline 
\cellcolor{verylow} defender  & \cellcolor{verylow} exclusions  & \cellcolor{verylow} paths  & \cellcolor{verylow} c  & \cellcolor{verylow} users  & \cellcolor{verylow} administrator \\ \hline 
\cellcolor{verylow} appdata  & \cellcolor{low} roaming  & \cellcolor{verylow} alfsvwjb  & \cellcolor{verylow} policies  & \cellcolor{verylow} microsoft  & \cellcolor{verylow} windows \\ \hline 
\cellcolor{verylow} nt  & \cellcolor{verylow} terminal  & \cellcolor{verylow} services  & \cellcolor{verylow} maxidletime  & \cellcolor{verylow} microsoft  & \cellcolor{verylow} windows \\ \hline 
\cellcolor{verylow} currentversion  & \cellcolor{low} explorer  & \cellcolor{high} advanced  & \cellcolor{veryhigh} showsuperhidden 
\end{tabular}
\caption{Attention for registry keys written shown for a malware sample that the model classified correctly as malicious. The cells are colored by how much attention each word received, with colors: \textcolor{veryhigh}{veryhigh}, \textcolor{high}{high}, \textcolor{med}{medium}, \textcolor{low}{low}, and \textcolor{verylow}{verylow}.}
\label{tab:attention_reg}
\end{table*}

\section{Datasets}
\begin{figure}
\begin{center}
\centerline{\includegraphics[width=0.8\linewidth]{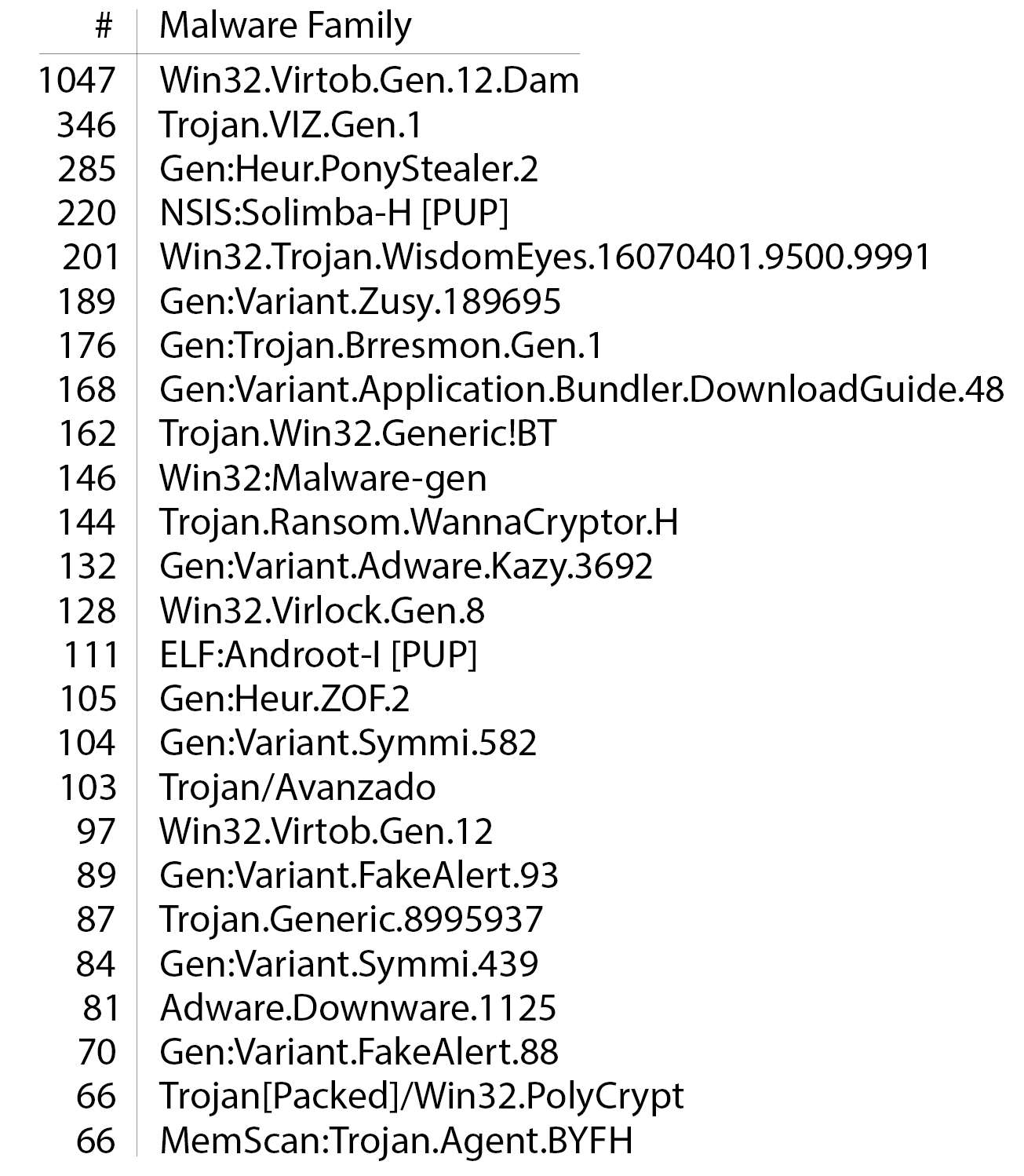}}
\caption{Top 25 Malware Families in \wilddataset and count(\#)}
\label{fig:malware_families}
\end{center}
\vskip -0.2in
\end{figure}

We have employed two different datasets for evaluating our research work. Dataset 1 (\wilddataset) is a set of 27,540 Windows x86 binaries with 13,760 benign and 13,760 malicious files. This private dataset was randomly selected from an original pool that was analyzed by the anti-malware vendor's sandbox in the US during the period from 2017-05-15 to 2017-09-19. This security vendor provides a sandbox that runs executables, and collects full analysis results that outline what the sample does while running inside an isolated operating system. Dataset 2 (\emberdataset) is a subset of the publicly available Ember dataset\cite{anderson2018ember}. It consists of 42,000 Windows x86 binaries with 21,000 benign and 21,000 malicious samples. \wilddataset has over a 1000 malware families, top 25 of which are shown in the Figure~\ref{fig:malware_families}.

\begin{figure}
\begin{center}
\centerline{\includegraphics[width=\linewidth]{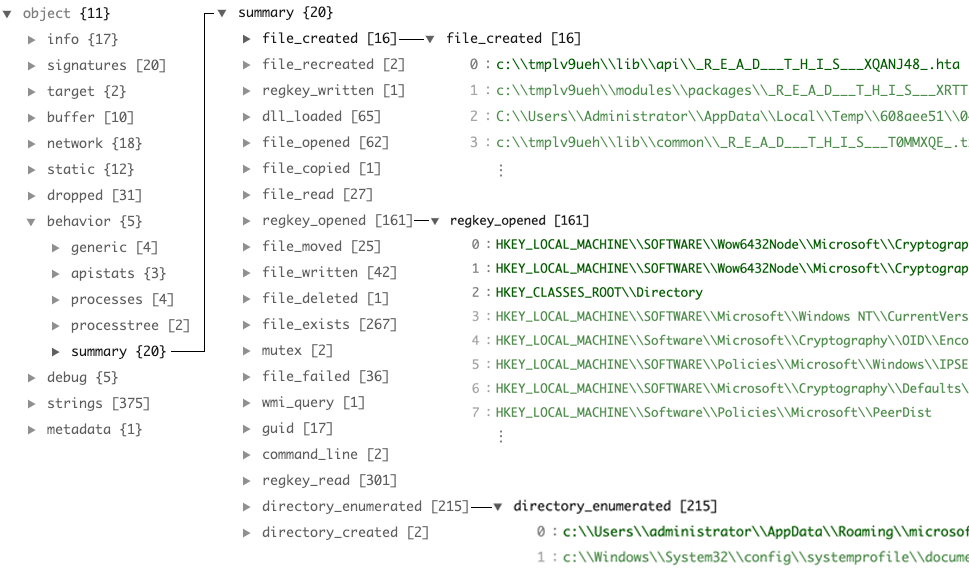}}
\caption{Cuckoo Sandbox Report format example}
\label{fig:cuckoo_report}
\end{center}
\vskip -0.2in
\end{figure}


Each binary is accompanied by two versions of detailed behavioral analysis reports in a JSON format. One behavioral report, \cuckooreport, is collected using the Cuckoo sandbox \cite{cuckoo}. Cuckoo is a publicly available sandbox which can be used to trace execution in a virtualized environment. It generates a JSON report of the actions taken by the binary during runtime. The other report, \lastlinereport, is collected using the sandbox from the security vendor, which also traces execution and collects information about the runtime execution of the executable.
This results in two datasets with two different report formats for a total of four different combinations of datasets and reports. 

\subsection{Comparison of Reports}
\label{sec:report_compare}
We ran an initial analysis of the two reports to understand their structure and the features they contain, which can be used for dynamic malware analysis. The format of one JSON report is shown in Figure~\ref{fig:cuckoo_report}. When examining the reports, we noticed that parts of the reports could be quite different for identical executables. The number of registry actions, file actions, and even the actual paths tended to differ. Differences show up because of differences in the sandbox and execution environment; even how long the executable is run influences the data in the reports. 

The feature names do not match up exactly, so we try to draw parallels in features from the respective sandboxes. For example, \lastlinereport references ``loaded\_libraries'', whereas \cuckooreport uses ``dll\_loaded'', but semantically they are the same. 
Due to subtle differences between the sandbox environments and formats, strings to do not match exactly. For example, when looking at registry keys, in the \lastlinereport we observe a key starting \texttt{hkcu\texttt{\char`\\}software\texttt{\char`\\}microsoft\texttt{\char`\\}windows}, while in Cuckoo it shows as \texttt{hkey\_current\_user\texttt{\char`\\}software\texttt{\char`\\}  microsoft\texttt{\char`\\} windows}.

This shows that the reports are similar but not identical, and thus, our model needs to be robust enough to handle the discrepancies. We will later evaluate how robust various approaches are to these differences between the two sandbox reports.

\begin{figure}
\begin{subfigure}
\centering
      \includegraphics[width=3.2in]{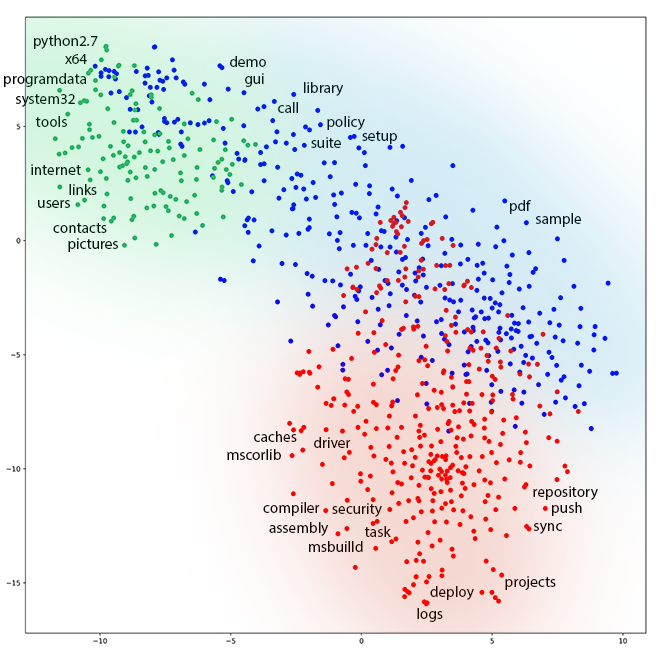}
      \caption{T-SNE visualization of file operations. This shows how the files used in file operations were clustered by the trained embedding layer. \textcolor{blue}{Blue} shows files common to benign files, \textcolor{red}{red} shows files common to malicious, and \textcolor{mediumseagreen}{green} shows files common to both.
      }
      \label{fig:tsne}
\end{subfigure}
\end{figure}


\section{Evaluation}
\label{sec:evaluation}

In this section, we evaluate \deepnet, which we described in Section ~\ref{sec:noengineering}. We compare \deepnet to approaches with  feature engineering, which are described in Section ~\ref{sec:featureengineering}.
We also compare \deepnet with MalDy~\cite{karbab2019maldy}. In particular, we attempt to answer the following research questions:

\begin{description}
    \item[RQ1:] Can deep learning methods without feature engineering identify malware from dynamic analysis reports as effectively as methods with feature engineering?
    \item[RQ2:] Does the application of advanced NLP techniques improve the results of malware detection on dynamic analysis reports compared to other deep learning models?
    \item[RQ3:] How robust are the various approaches to being applied to different datasets and sandbox/reports?
    \item[RQ4:] Which parts of the report does \deepnet learn to use in detecting malware?
\end{description}

The first two research questions will help us to evaluate \deepnet's malware detection capabilities, and compare our approach to other deep learning models. The third question is useful in order to understand if our approach is learning robust features that apply to other environments.
Finally, the last question is chosen to explore which parts of the report are used by \deepnet. This can help in determining if the approach is learning useful features.

\subsection{Experiment Design}
Here we will evaluate the performance of \deepnet and each of the comparison models described in the previous section. 
All the evaluated models are listed in Table~\ref{tab:modelrecap} with a brief recap on their approach.
We trained each model on reports from the \lastlinereport on executables from the \emberdataset. We chose the \emberdataset for our training because it is the larger dataset and would provide more samples for training.
We performed the classical k-fold cross validation (where $k = 10$) to test the models. That is, we divide the dataset $D$ into $D_1,D_2,...,D_k$. We spare $D_i$ for $1<i<k$ for testing, and use the remaining $k-1$ folds for training. This process is repeated $k$ times to get accurate validation results.
We trained all our models by minimizing the cross-entropy error.

After this, each model is then tested on the \wilddataset to evaluate its ability to generalize to a new dataset.
Each model is also tested using the \cuckooreport on the same samples in the \emberdataset to evaluate the robustness of the model to a different report from a different sandbox. 
These tests should show if the model is learning features that generalize well and are not specific to the particular report or dataset. 

Additionally, we compared against \maldy \cite{karbab2019maldy}. This is a state of the art model using XG-Boost and feature hashing. We implemented it for our tests as we could not obtain the source code.

\begin{table*}
\begin{center}
\begin{tabular}{|l|l|l|l|l|l|l|}
\hline
Model & Accuracy & Precision & Recall & F1-Score & 
\begin{tabular}[c]{@{}l@{}} Accuracy on \\ \wilddataset \end{tabular} &
\begin{tabular}[c]{@{}l@{}} Accuracy on \\ \cuckooreport \end{tabular}  \\ \hline

\maldy \cite{karbab2019maldy}  & .8923 & .850 & .830 & .885 & .55 & .40 \\ \hline
\countsmodel & .935 & .891 & .941 & .92 & .778 & $-$ \\ \hline
\ensemblemodel & .980 & .986 & .975 & .980 & .843 & .780 \\ \hline
\rawmodel & .914 & .948 & .906 & .927 & .698 & .627 \\ \hline
\textbf{\deepnet} & .968 & .967 & .966 & .968 & .870 & .867 \\ \hline
\end{tabular}
\caption{This table shows how the models performed. The models were all trained on reports from \lastlinereport on the \emberdataset. They are tested using K-Fold validation and tested on the other dataset and report format. The first row, \maldy, is a state of the art model from \cite{karbab2019maldy} that we compare with. The next two rows are feature-engineering based approaches.}
\label{tab:results}
\end{center}
\end{table*}


\begin{table}
\begin{center}
\begin{tabular}{|l|l|l|l|}
\hline
\begin{tabular}[c]{@{}l@{}} Individual \\ Feature \end{tabular} &
Accuracy &
\begin{tabular}[c]{@{}l@{}} Accuracy on \\ \wilddataset \end{tabular} &
\begin{tabular}[c]{@{}l@{}} Accuracy on \\ \cuckooreport \end{tabular}  \\ \hline

Registry & .944 & .801 & .767 \\ \hline
File & .978 & .842 & .771 \\ \hline
API  & .865 & .776 & .510 \\ \hline
Mutex & .808 & .733 & .677 \\ \hline
DLLs & .960 & .810 & .701 \\ \hline

\end{tabular}
\caption{This table shows how the model performed on each of the five individual features. This table shows the validation accuracy as well as the accuracy when given a different dataset and different report format. 
}
\label{tab:featureresults}
\end{center}
\end{table}

\subsection{Results}
 Table~\ref{tab:results} shows the results of each of the models we trained. Looking at the results, we see that \deepnet performs very well, showing the best accuracy when applied on both a new dataset and on a new sandbox, getting 87.0\% and 86.7\% respectively. The ensemble classifier outperformed it slightly in terms of validation accuracy, but it wasn't nearly as robust to differences in datasets or sandboxes. This result allows us to answer our first research question.

\begin{mdframed}
\textbf{Answer for RQ1}: \deepnet, a deep learning method without feature engineering performs about as well in validation accuracy as our best model with feature engineering. It also showed better results than the feature engineered models when tested on another dataset or report format.
\end{mdframed}

\subsection{Individual Features}
\label{sec:individresults}
We also trained and tested each of the five individual features we extracted using the same CNN+BiLSTM+Attention model design that we use in \deepnet. These individual models are also used to compose the \ensemblemodel. The results of each model are shown in Table~\ref{tab:featureresults}. We observed that file actions perform the best of any of our features. It also generalized fairly, showing good results on the other dataset. Note that APIs were represented significantly differently between the two sandboxes, which explains its low score on \cuckooreport. 

We visualized the trained word embeddings for file actions to see if the word embeddings are creating good clusters. For this we use a T-SNE plot \cite{maaten2008visualizing}, which is shown in Figure~\ref{fig:tsne}. In the T-SNE we see clusters of similar files that the model learned. We also see that the two clusters for files seen in benign and malicious files only had a bit of overlap.

\subsection{Attention}
When we look at the results in Table~\ref{tab:results}, we find that the \rawmodel had a lower validation accuracy than \deepnet, and performed much worse when tested on a different dataset. This implies that the \rawmodel might not be using very general features, whereas \deepnet appears to be learning features that generalize better.
In this section, we explore what the two models appear to be paying attention to, or in other words, which parts of the reports contribute most to the classification decisions by the models. This can be used to both understand the model better, and it can also be used by security researchers to identify possible features that they can use in other analyses.

\begin{figure}
\begin{subfigure}
\centering
      \fbox{
      \includegraphics[width=0.75\linewidth]{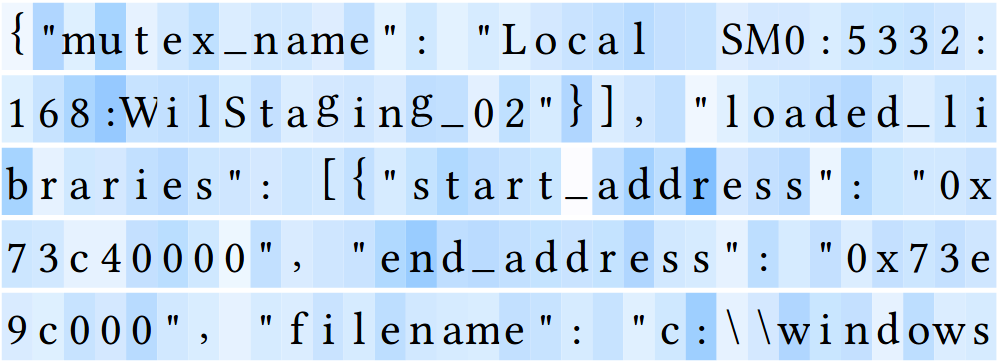}
      }
      \caption{The most salient area of sample 234 highlighted by the \rawmodel at the byte level showing the \rawmodel giving a lot of importance to seemingly random parts such as part of ''address`` and ''end``. Sample 234 was misclassified by the \rawmodel as benign when it is malicious.}
      \label{fig:salient}
\end{subfigure}
\end{figure}

Firstly, we examine the \rawmodel. To do this, we use a concept called saliency, introduced by Simonyan et al. \cite{simonyan2013deep}. Saliency uses the gradient of the output with respect to the input to determine which areas of input will most affect the output of a neural net. This will allow us to highlight, for a particular sample, the regions of bytes that contribute most to its classification. When applying saliency on the \rawmodel, we see that the model frequently pays the most attention to hashes (SHA-1 hashes, MD5 hashes), DLLs, file names, and library addresses. An example of the most highlighted area of one sample is shown in Figure~\ref{fig:salient}. Although SHA-1 hashes are frequently important to the model's classifications results, they are not a feature that generalizes well. Any byte changed can cause these features to be wrong. Additionally, intuitively, we know that library addresses are likely to change, and not the best feature either.

\begin{figure}
\begin{subfigure}
\centering
      \fbox{
      \includegraphics[width=0.8\linewidth]{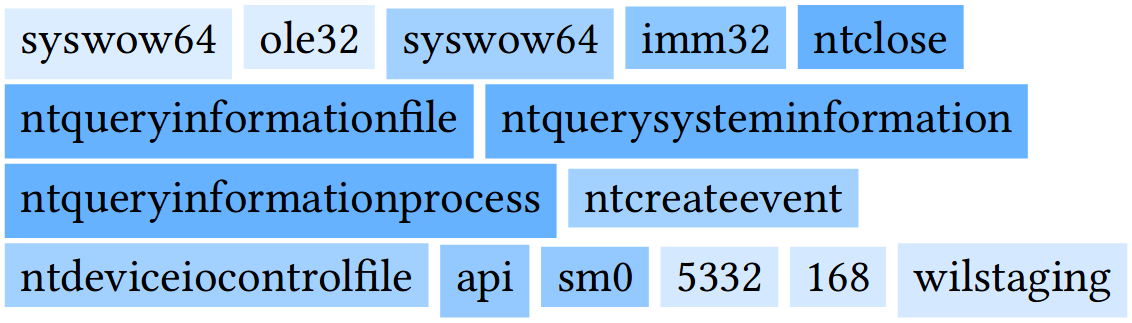}
      }
      \caption{The highest attention areas of sample 234 highlighted by \deepnet. \deepnet paid the most attention to the api calls in this example. \deepnet was able to correctly classify it as malicious.}
      \label{fig:attention}
\end{subfigure}
\end{figure}

Then, we examine our approach, \deepnet, which uses NLP techniques to learn on whole words rather than bytes. For this model, we have an Attention layer, explained in the Section~\ref{sec:attention}, which allows the model to learn what it should focus on, and allows us to interpret which words/phrases received the most attention directly. By examining the attention activations, we can also see what the model is paying attention. An example is shown in Figure~\ref{fig:attention}, which shows our method giving the most attention to a few API calls. Overall, we found that \deepnet focused on parts of the document that seemed much more general. For example, in the first couple of samples it focused on words such as ``ntqueryinformationprocess'', ``ntreadvirtualmemory'', ``virustotal'',  ``programs'', ``startup''. These intuitively make sense as valuable features. The first couple indicates that the process is attempting to interact with other running processes. Then, of course, ``startup programs'' shows that the executable might be trying to set a new process to autostart.

Our approach (\deepnet) was able to pick features that look better intuitively, and it showed that its results generalized well to other datasets. This seems to be a result of applying document classification techniques from NLP to our model. Our approach looks at whole words rather than bytes, and its model learns better which words and phrases are indicative of malicious behavior when compared with the raw model, which focused on more arbitrary things such as sequences of bytes in SHA1 or MD5 hashes. 

\begin{table}
\begin{center}
\begin{tabular}{|l|l|}
\hline
Feature & 
\begin{tabular}[c]{@{}l@{}} Number of times \\ in top attention \end{tabular} \\ \hline
File & 831 \\ \hline
API & 305 \\ \hline
Network  & 107 \\ \hline
DLL Loads & 95 \\ \hline
Mutex & 72 \\ \hline
Registry & 20 \\ \hline

\end{tabular}
\caption{This table shows how many times each feature appeared as one of the 10 most important words according to the attention score for 100 samples.
}
\label{tab:topattention}
\end{center}
\end{table}

We also counted which features show up most frequently with the highest attention score. More specifically, for a subset of malicious samples we took the words with the most attention and determined to which feature they applied. These results are shown in Table~\ref{tab:topattention}. We see that file operations were most commonly the most highly paid attention to, followed by API calls, then network and DLL loads. Interestingly, file operations were also the best performing individual feature. This means that the feature it paid the most attention to was also the best performing on its own.

\begin{mdframed}
\textbf{Answer for RQ2}: NLP techniques for document classification can be effectively applied on reports to perform malware detection and show much better results than our \rawmodel neural network.
\end{mdframed}

\begin{mdframed}
\textbf{Answer for RQ4}: \deepnet appears to be learning to use the best combinations of features. Specifically, it pays more attention to the file operations performed by the malware, as well as the API calls.
\end{mdframed}

\subsection{Robustness}
We found that all the approaches had lower accuracies when tested on a dataset or sandbox report that they were not trained on. The report format has many differences that should account for the drop in accuracy. However, the lower accuracy on \wilddataset implies that we are learning features that do not generalize to all executables. This could be a problem due to deficiencies in the training set (not having as wide a breadth of samples as we need). Also, it could be that the model is still learning some specific features that do not generalize as well.
\deepnet showed the best robustness to the report formats. Also, the other NLP-based deep learning approaches (the individual models and \ensemblemodel) showed decent robustness, implying that the NLP techniques give us more general features than our raw bytes approach. 

\begin{mdframed}
\textbf{Answer for RQ3}: \deepnet is the most robust of the models we tried, showing the highest accuracy on another dataset and on another report format. On the other hand, our raw bytes model was poor at classification across datasets and reports. This implies that the features learned using text classification approaches were more general.
\end{mdframed}

\subsection{Unseen Malware Family}
Another experiment we performed was to remove one malware family, train on the remaining data, and then evaluate \deepnet's classification accuracy when tested on that family. We removed all samples that were identified as the family Trojan.Viz.Gen.1, using VirusTotal. During evaluation \deepnet still correctly classified all 346 samples of that family as malware.

\subsection{Performance}
We run our experiments on Nvidia Titan RTX and Xeon Gold 6252 processor. Our training process took $19.47$ $milli seconds / sample$ and detection process took $8.21$ $milli seconds/ sample$. The data-cleaning/preprocessing runs at a rate of $6.16$ $milliseconds/ sample$.

\section{Discussion and Future Work}
One limitation of \deepnet is that it performs classification based on behavior seen in dynamic analyses. This means that it is not as effective as a preventative measure. However, its results still are useful for identifying malware to generate signatures or catch that an infection has occurred. Also, \deepnet showed that it was able to detect a previously unseen family, indicating that it can be used even on malware that has not been analyzed before. This result could be further explored to understand the correlations between different malware families.

Also, \deepnet relies on accurate and broad training data. Future work would be to try and make it more resilient to the quality of the training data, as well as exploring different execution environments to discover if one provides better results for it. Other directions for future work include exploring different models to improve the results. For example, image recognition \cite{long2015learning}, has recently shown promising results. These models are based on ones that were known to perform well on image recognition tasks, and can also sometimes use transfer learning, using the training already done on images.

\subsection{Adversarial Learning}
Some recent works try to evade the detection of machine learning based malware classifiers by adversarial learning. Their experiments show that it is possible to generate adversarial samples based on a trained machine learning classifier. The core of adversarial sample crafting is to find a small perturbation to feature vectors \(X\) of the original malware sample to change the classification results \(F\) to benign. Formally, they compute the gradient of \(F\) with respect to \(X\) to estimate the direction in which a perturbation in \(X\) would maximally change \(F's\) output. The earliest work of this topic came from Nguyen et al. \cite{nguyen2015deep} who found that a slight change in the image could trick the image classifier, and then it has been introduced into computer security in recent years to attack security systems that rely on machine learning models. 


Robustness against adversarial attacks provided is an essential design characteristic. Our future work will include making our proposed model more robust against such attacks.

\section{Related Work}
There has been much work on using machine learning and NLP models to classify malware, all bringing new strengths and trade-offs to the table. There were initially many machine learning models being used, such as Support Vector Machines (SVM), Decision Trees (DT), Random Forest, and K-Nearest Neighbor (KNN) amongst others. Recently, neural networks have been prevalent for detecting and classifying malware. 

Saxe \etal~\cite{saxe2015deep} proposed a method to distinguish malware from benign software based on a neural network, deep learning approach. Their system  uses four different types of complementary static features from benign and malicious binaries. These features are entropy histogram features, PE import features, string 2D histogram features, and PE metadata features. However, this work primarily focused on feature engineering and static analysis. 
Fan \etal~\cite{fan2016malicious} created a sequence mining algorithm that discovers consecutive malicious patterns using All-Nearest-Neighbor classifier. They used feature engineering to extract instruction sequences from the PE files as the preliminary features. This approach used static features and required domain knowledge for feature extraction.

Zheng \etal~\cite{zhang2019maldc} creates a behavior chain that aids in its detection method. The method monitors behavior points based on API calls and then uses the respective calling sequence at runtime to construct a behavior chain. The system uses long short-term memory (LSTM) to detect maliciousness from the created chains. 
Salehi \etal~\cite{salehi2017maar} used a monitored environment where the arguments and return values of every API call are recorded as (API, variables, return value) tri-tuples. The authors found a selective and discriminative set of these features and used SVM algorithm for classification. Other methods also focused on using API sequence calls for dynamic malware classification \cite{shankarapani2011malware, peiravian2013machine, sami2010malware}. However, restricting to only one type of feature limits the vast feature space that can be used to represent different types of malware behaviors.   

MalInsight ~\cite{han2019malinsight} takes a different approach, and profiles malware based on basic structure, low-level behavior, and high-level behavior. A feature space is built on three core profiles, namely, the structural features, how the binary interacts with the OS, and operations on files, registry, and network. Han et al. pick select features from the Cuckoo sandbox and train on those. Unlike our experiments in Section~\ref{sec:individresults}, they do not use neural networks and they do not use the whole report as \deepnet does, instead requiring feature selection.

Another approach is based on early stage recognition \cite{rhode2018early}. They implemented a recurrent neural network that takes as input a short feature set of file activity. They can predict whether or not a file is malicious using the first few seconds of file execution. Their intuition is that malicious activity surfaces rapidly once the malicious file begins execution. Their approach differs from ours in that it only uses file activity, and aims to detect malware in real time. We use entire dynamic analysis reports which provide a more complete picture but requires analyzing it in a sandbox.

Zhong \etal~\cite{zhong2019multi} suggested that a single deep learning model was insufficient and created a Multi-Level Deep Learning System. They first partition the data using static and dynamic features, then create a convolutional model which learns to classify each cluster, and combine the clusters to create their final model.
They use a feature extraction phase to extract static and dynamic features, whereas \deepnet simply learns on the report of dynamic behavior and not extracted features.

Another approach that performs classification on behavior reports is MalDy, which uses a bag of words approach, with models combined in an ensemble \cite{karbab2019maldy}. A limitation of a bag of words approach is that it does not take into account the context, just the frequencies with which words appear \cite{paltoglou2013more}. Additionally, BoW produces feature vectors which are fairly sparse. On the other hand, our system, which uses word embeddings does not have this limitation, and produces more dense, low dimensional feature vectors. In our evaluation (results in Table~\ref{tab:results}), we show that \deepnet gives a higher validation accuracy and a much stronger ability to generalize, both to other datasets and to other report formats than MalDy.

\section{Conclusions}
This paper introduced \deepnet, a robust malware detection tool that can successfully identify malicious files based on their run-time behavior without feature engineering. It is based on techniques borrowed from the field of document classification and applied to dynamic analysis reports. Based on our evaluation results, we conclude that \deepnet outperforms similar approaches for malware classification. Our work not only focused on eliminating the need for feature engineering, but also explains the relation of the classification process with respect to different auto-detected features. 
The fact that \deepnet can retain a high detection accuracy when tested on samples from another dataset and on an unknown report format shows that our model promises robust real-world applicability. 


\begin{acks}
This research is based on research sponsored by a gift from Intel for the investigation of machine learning for malware analysis, by the National Science Foundation grant \#CNS-1704253, and by DARPA under agreement number \#FA8750-19-C-0003. The U.S. Government is authorized to reproduce and distribute reprints for Governmental purposes notwithstanding any copyright notation thereon. The views and conclusions contained herein are those of the authors and should not be interpreted as necessarily representing the official policies or endorsements, either expressed or implied, of DARPA or the U.S. Government.

We would also like to thank Lastline for providing data that made this research possible. 
\end{acks}

\bibliographystyle{abbrvnat}
\bibliography{main}

\begin{thebibliography}{45}
\providecommand{\natexlab}[1]{#1}
\providecommand{\url}[1]{\texttt{#1}}
\expandafter\ifx\csname urlstyle\endcsname\relax
  \providecommand{\doi}[1]{doi: #1}\else
  \providecommand{\doi}{doi: \begingroup \urlstyle{rm}\Url}\fi

\bibitem[cuc()]{cuckoo}
Cuckoo, automated malware analysis.
\newblock \url{https://cuckoosandbox.org/}.

\bibitem[Anderson and Roth(2018)]{anderson2018ember}
H.~S. Anderson and P.~Roth.
\newblock Ember: an open dataset for training static pe malware machine
  learning models.
\newblock \emph{arXiv preprint arXiv:1804.04637}, 2018.

\bibitem[Brosch and Morgenstern(2006)]{brosch2006runtime}
T.~Brosch and M.~Morgenstern.
\newblock {Runtime Packers: The Hidden Problem?}
\newblock \emph{Black Hat USA}, 2006.

\bibitem[Fan et~al.(2016)Fan, Ye, and Chen]{fan2016malicious}
Y.~Fan, Y.~Ye, and L.~Chen.
\newblock Malicious sequential pattern mining for automatic malware detection.
\newblock \emph{Expert Systems with Applications}, 52:\penalty0 16--25, 2016.

\bibitem[Garfinkel et~al.(2007)Garfinkel, Adams, Warfield, and
  Franklin]{garfinkel2007compatibility}
T.~Garfinkel, K.~Adams, A.~Warfield, and J.~Franklin.
\newblock Compatibility is not transparency: Vmm detection myths and realities.
\newblock In \emph{HotOS}, 2007.

\bibitem[Grosse et~al.(2017)Grosse, Papernot, Manoharan, Backes, and
  McDaniel]{grosse2017adversarial}
K.~Grosse, N.~Papernot, P.~Manoharan, M.~Backes, and P.~McDaniel.
\newblock Adversarial examples for malware detection.
\newblock In \emph{European Symposium on Research in Computer Security}, pages
  62--79. Springer, 2017.

\bibitem[Han et~al.(2019)Han, Xue, Wang, Liu, and Kong]{han2019malinsight}
W.~Han, J.~Xue, Y.~Wang, Z.~Liu, and Z.~Kong.
\newblock Malinsight: A systematic profiling based malware detection framework.
\newblock \emph{Journal of Network and Computer Applications}, 125:\penalty0
  236--250, 2019.

\bibitem[Hochreiter(1998)]{hochreiter1998vanishing}
S.~Hochreiter.
\newblock The vanishing gradient problem during learning recurrent neural nets
  and problem solutions.
\newblock \emph{International Journal of Uncertainty, Fuzziness and
  Knowledge-Based Systems}, 6\penalty0 (02):\penalty0 107--116, 1998.

\bibitem[Hochreiter and Schmidhuber(1997)]{hochreiter1997long}
S.~Hochreiter and J.~Schmidhuber.
\newblock Long short-term memory.
\newblock \emph{Neural computation}, 9\penalty0 (8):\penalty0 1735--1780, 1997.

\bibitem[{Kanchan Sarkar}(2017)]{trendsinnlp}
{Kanchan Sarkar}.
\newblock Recent trends in natural language processing using deep learning,
  2017.
\newblock
  \url{https://medium.com/@kanchansarkar/recent-trends-in-natural-language-processing-using-deep-learning-a1469fbd2ef}.

\bibitem[Karbab and Debbabi(2019)]{karbab2019maldy}
E.~B. Karbab and M.~Debbabi.
\newblock Maldy: Portable, data-driven malware detection using natural language
  processing and machine learning techniques on behavioral analysis reports.
\newblock \emph{Digital Investigation}, 28:\penalty0 S77--S87, 2019.

\bibitem[Kharaz et~al.(2016)Kharaz, Arshad, Mulliner, Robertson, and
  Kirda]{kharaz2016unveil}
A.~Kharaz, S.~Arshad, C.~Mulliner, W.~Robertson, and E.~Kirda.
\newblock $\{$UNVEIL$\}$: A large-scale, automated approach to detecting
  ransomware.
\newblock In \emph{25th $\{$USENIX$\}$ Security Symposium ($\{$USENIX$\}$
  Security 16)}, pages 757--772, 2016.

\bibitem[Kolbitsch et~al.(2009)Kolbitsch, Comparetti, Kruegel, Kirda, Zhou, and
  Wang]{kolbitsch2009effective}
C.~Kolbitsch, P.~M. Comparetti, C.~Kruegel, E.~Kirda, X.-y. Zhou, and X.~Wang.
\newblock Effective and efficient malware detection at the end host.
\newblock In \emph{USENIX security symposium}, volume~4, pages 351--366, 2009.

\bibitem[Kolosnjaji et~al.(2016)Kolosnjaji, Zarras, Webster, and
  Eckert]{kolosnjaji2016deep}
B.~Kolosnjaji, A.~Zarras, G.~Webster, and C.~Eckert.
\newblock Deep learning for classification of malware system call sequences.
\newblock In \emph{Australasian Joint Conference on Artificial Intelligence},
  pages 137--149. Springer, 2016.

\bibitem[Lindorfer et~al.(2011)Lindorfer, Kolbitsch, and
  Comparetti]{lindorfer2011detecting}
M.~Lindorfer, C.~Kolbitsch, and P.~M. Comparetti.
\newblock Detecting environment-sensitive malware.
\newblock In \emph{International Workshop on Recent Advances in Intrusion
  Detection}, pages 338--357. Springer, 2011.

\bibitem[Long et~al.(2015)Long, Cao, Wang, and Jordan]{long2015learning}
M.~Long, Y.~Cao, J.~Wang, and M.~I. Jordan.
\newblock Learning transferable features with deep adaptation networks.
\newblock \emph{arXiv preprint arXiv:1502.02791}, 2015.

\bibitem[Maaten and Hinton(2008)]{maaten2008visualizing}
L.~v.~d. Maaten and G.~Hinton.
\newblock Visualizing data using t-sne.
\newblock \emph{Journal of machine learning research}, 9\penalty0
  (Nov):\penalty0 2579--2605, 2008.

\bibitem[Mikolov et~al.(2013)Mikolov, Sutskever, Chen, Corrado, and
  Dean]{mikolov2013distributed}
T.~Mikolov, I.~Sutskever, K.~Chen, G.~S. Corrado, and J.~Dean.
\newblock Distributed representations of words and phrases and their
  compositionality.
\newblock In \emph{Advances in neural information processing systems}, pages
  3111--3119, 2013.

\bibitem[Moser et~al.(2007)Moser, Kruegel, and Kirda]{moser2007limits}
A.~Moser, C.~Kruegel, and E.~Kirda.
\newblock Limits of static analysis for malware detection.
\newblock In \emph{Twenty-Third Annual Computer Security Applications
  Conference (ACSAC 2007)}, pages 421--430. IEEE, 2007.

\bibitem[Nguyen et~al.(2015)Nguyen, Yosinski, and Clune]{nguyen2015deep}
A.~Nguyen, J.~Yosinski, and J.~Clune.
\newblock Deep neural networks are easily fooled: High confidence predictions
  for unrecognizable images.
\newblock In \emph{Proceedings of the IEEE Conference on Computer Vision and
  Pattern Recognition}, pages 427--436, 2015.

\bibitem[Oktavianto and Muhardianto(2013)]{oktavianto2013cuckoo}
D.~Oktavianto and I.~Muhardianto.
\newblock \emph{Cuckoo malware analysis}.
\newblock Packt Publishing Ltd, 2013.

\bibitem[Paltoglou and Thelwall(2013)]{paltoglou2013more}
G.~Paltoglou and M.~Thelwall.
\newblock More than bag-of-words: Sentence-based document representation for
  sentiment analysis.
\newblock In \emph{Proceedings of the International Conference Recent Advances
  in Natural Language Processing RANLP 2013}, pages 546--552, 2013.

\bibitem[Peiravian and Zhu(2013)]{peiravian2013machine}
N.~Peiravian and X.~Zhu.
\newblock Machine learning for android malware detection using permission and
  api calls.
\newblock In \emph{2013 IEEE 25th international conference on tools with
  artificial intelligence}, pages 300--305. IEEE, 2013.

\bibitem[Perdisci et~al.(2008)Perdisci, Lanzi, and Lee]{perdisci2008mcboost}
R.~Perdisci, A.~Lanzi, and W.~Lee.
\newblock Mcboost: Boosting scalability in malware collection and analysis
  using statistical classification of executables.
\newblock In \emph{2008 Annual Computer Security Applications Conference
  (ACSAC)}, pages 301--310. IEEE, 2008.

\bibitem[Raff et~al.(2017)Raff, Sylvester, and Nicholas]{raff2017learning}
E.~Raff, J.~Sylvester, and C.~Nicholas.
\newblock Learning the pe header, malware detection with minimal domain
  knowledge.
\newblock In \emph{Proceedings of the 10th ACM Workshop on Artificial
  Intelligence and Security}, pages 121--132. ACM, 2017.

\bibitem[Raff et~al.(2018)Raff, Barker, Sylvester, Brandon, Catanzaro, and
  Nicholas]{raff2018malware}
E.~Raff, J.~Barker, J.~Sylvester, R.~Brandon, B.~Catanzaro, and C.~K. Nicholas.
\newblock Malware detection by eating a whole exe.
\newblock In \emph{Workshops at the Thirty-Second AAAI Conference on Artificial
  Intelligence}, 2018.

\bibitem[Raffetseder et~al.(2007)Raffetseder, Kruegel, and
  Kirda]{raffetseder2007detecting}
T.~Raffetseder, C.~Kruegel, and E.~Kirda.
\newblock Detecting system emulators.
\newblock In \emph{International Conference on Information Security}, pages
  1--18. Springer, 2007.

\bibitem[Rahbarinia et~al.(2017)Rahbarinia, Balduzzi, and
  Perdisci]{rahbarinia2017exploring}
B.~Rahbarinia, M.~Balduzzi, and R.~Perdisci.
\newblock Exploring the long tail of (malicious) software downloads.
\newblock In \emph{2017 47th Annual IEEE/IFIP International Conference on
  Dependable Systems and Networks (DSN)}, pages 391--402. IEEE, 2017.

\bibitem[Rhode et~al.(2018)Rhode, Burnap, and Jones]{rhode2018early}
M.~Rhode, P.~Burnap, and K.~Jones.
\newblock Early-stage malware prediction using recurrent neural networks.
\newblock \emph{computers \& security}, 77:\penalty0 578--594, 2018.

\bibitem[Rossow et~al.(2012)Rossow, Dietrich, Grier, Kreibich, Paxson,
  Pohlmann, Bos, and Van~Steen]{rossow2012prudent}
C.~Rossow, C.~J. Dietrich, C.~Grier, C.~Kreibich, V.~Paxson, N.~Pohlmann,
  H.~Bos, and M.~Van~Steen.
\newblock Prudent practices for designing malware experiments: Status quo and
  outlook.
\newblock In \emph{2012 IEEE Symposium on Security and Privacy}, pages 65--79.
  IEEE, 2012.

\bibitem[Salehi et~al.(2017)Salehi, Sami, and Ghiasi]{salehi2017maar}
Z.~Salehi, A.~Sami, and M.~Ghiasi.
\newblock Maar: Robust features to detect malicious activity based on api
  calls, their arguments and return values.
\newblock \emph{Engineering Applications of Artificial Intelligence},
  59:\penalty0 93--102, 2017.

\bibitem[Sami et~al.(2010)Sami, Yadegari, Rahimi, Peiravian, Hashemi, and
  Hamze]{sami2010malware}
A.~Sami, B.~Yadegari, H.~Rahimi, N.~Peiravian, S.~Hashemi, and A.~Hamze.
\newblock Malware detection based on mining api calls.
\newblock In \emph{Proceedings of the 2010 ACM symposium on applied computing},
  pages 1020--1025. ACM, 2010.

\bibitem[Saxe and Berlin(2015)]{saxe2015deep}
J.~Saxe and K.~Berlin.
\newblock Deep neural network based malware detection using two dimensional
  binary program features.
\newblock In \emph{Malicious and Unwanted Software (MALWARE), 2015 10th
  International Conference on}, pages 11--20. IEEE, 2015.

\bibitem[Sgandurra et~al.(2016)Sgandurra, Mu{\~n}oz-Gonz{\'a}lez, Mohsen, and
  Lupu]{sgandurra2016automated}
D.~Sgandurra, L.~Mu{\~n}oz-Gonz{\'a}lez, R.~Mohsen, and E.~C. Lupu.
\newblock Automated dynamic analysis of ransomware: Benefits, limitations and
  use for detection.
\newblock \emph{arXiv preprint arXiv:1609.03020}, 2016.

\bibitem[Shankarapani et~al.(2011)Shankarapani, Ramamoorthy, Movva, and
  Mukkamala]{shankarapani2011malware}
M.~K. Shankarapani, S.~Ramamoorthy, R.~S. Movva, and S.~Mukkamala.
\newblock Malware detection using assembly and api call sequences.
\newblock \emph{Journal in computer virology}, 7\penalty0 (2):\penalty0
  107--119, 2011.

\bibitem[Simonyan and Zisserman(2014)]{simonyan2014very}
K.~Simonyan and A.~Zisserman.
\newblock Very deep convolutional networks for large-scale image recognition.
\newblock \emph{arXiv preprint arXiv:1409.1556}, 2014.

\bibitem[Simonyan et~al.(2013)Simonyan, Vedaldi, and
  Zisserman]{simonyan2013deep}
K.~Simonyan, A.~Vedaldi, and A.~Zisserman.
\newblock Deep inside convolutional networks: Visualising image classification
  models and saliency maps.
\newblock \emph{arXiv preprint arXiv:1312.6034}, 2013.

\bibitem[Ucci et~al.(2017)Ucci, Aniello, and Baldoni]{ucci2017survey}
D.~Ucci, L.~Aniello, and R.~Baldoni.
\newblock Survey on the usage of machine learning techniques for malware
  analysis.
\newblock \emph{arXiv preprint arXiv:1710.08189}, 2017.

\bibitem[Vaswani et~al.(2017)Vaswani, Shazeer, Parmar, Uszkoreit, Jones, Gomez,
  Kaiser, and Polosukhin]{vaswani2017attention}
A.~Vaswani, N.~Shazeer, N.~Parmar, J.~Uszkoreit, L.~Jones, A.~N. Gomez,
  {\L}.~Kaiser, and I.~Polosukhin.
\newblock Attention is all you need.
\newblock In \emph{Advances in neural information processing systems}, pages
  5998--6008, 2017.

\bibitem[VirusTotal()]{vt-comparative-analyses}
VirusTotal.
\newblock Av comparative analyses.
\newblock
  \url{https://blog.virustotal.com/2012/08/av-comparative-analyses-marketing-and.html}.
\newblock (Accessed: 2019-3-31).

\bibitem[Yang et~al.(2016)Yang, Yang, Dyer, He, Smola, and
  Hovy]{yang2016hierarchical}
Z.~Yang, D.~Yang, C.~Dyer, X.~He, A.~Smola, and E.~Hovy.
\newblock Hierarchical attention networks for document classification.
\newblock In \emph{Proceedings of the 2016 Conference of the North American
  Chapter of the Association for Computational Linguistics: Human Language
  Technologies}, pages 1480--1489, 2016.

\bibitem[Zeltser()]{malwaremutex}
L.~Zeltser.
\newblock How malware generates mutex names to evade detection.
\newblock
  \url{https://isc.sans.edu/diary/How+Malware+Generates+Mutex+Names+to+Evade+Detection/19429/}.
\newblock (Accessed: 2019-5-31).

\bibitem[Zhang et~al.(2019)Zhang, Zhang, Lv, Sangaiah, Huang, and
  Chilamkurti]{zhang2019maldc}
H.~Zhang, W.~Zhang, Z.~Lv, A.~K. Sangaiah, T.~Huang, and N.~Chilamkurti.
\newblock Maldc: a depth detection method for malware based on behavior chains.
\newblock \emph{World Wide Web}, pages 1--20, 2019.

\bibitem[Zhong and Gu(2019)]{zhong2019multi}
W.~Zhong and F.~Gu.
\newblock A multi-level deep learning system for malware detection.
\newblock \emph{Expert Systems with Applications}, 2019.

\bibitem[Zhou et~al.(2015)Zhou, Sun, Liu, and Lau]{zhou2015c}
C.~Zhou, C.~Sun, Z.~Liu, and F.~Lau.
\newblock A c-lstm neural network for text classification.
\newblock \emph{arXiv preprint arXiv:1511.08630}, 2015.

\end{thebibliography}

\end{document}